\documentclass[a4paper,11pt]{article}
\usepackage{pos}

\usepackage{graphicx}
\usepackage{tabularx}
\usepackage{slashed}
\usepackage{mathtools}

\def\bea{\begin{eqnarray}}
\def\eea{\end{eqnarray}}
\def\eps{\epsilon}
\def\be{\begin{equation}}
\def\ee{\end{equation}}

\title{Update of the Standard-Model prediction for $\bar B \to X_s \gamma$}
\author*[a]{Tobias Huber}
\affiliation[a]{Theoretische Physik 1, Center for Particle Physics Siegen (CPPS), Universit\"at Siegen, \\
Walter-Flex-Str.~3, 57068 Siegen, Germany}

\emailAdd{huber@tp1.physik.uni-siegen.de}

\abstract{We report on two recent calculations on the inclusive radiative
decay $\bar B\to X_s \gamma$, notably multi-parton contributions at NLO and
the $Q_{1,2}-Q_7$ interference at NNLO for the physical value of the charm-quark mass.
The former calculation formally completes $\bar{B} \rightarrow X_s \gamma$ at NLO in QCD at leading power, the latter removes a long-standing $\pm 3\%$ uncertainty arising from interpolation in $m_c$. The updated Standard-Model prediction
for the CP- and isospin-averaged branching ratio reads
${\mathcal B}_{s \gamma}^{\rm SM} = (3.54 \pm 0.14)\times 10^{-4}$
for photon energies $E_\gamma > 1.6\,{\rm GeV}$ in the $B$-meson rest frame.
This value is in good agreement with the current experimental average
${\mathcal B}_{s \gamma}^{\rm exp} = (3.49 \pm 0.19)\times 10^{-4}$.}

\FullConference{Loops and Legs in Quantum Field Theory (LL2026)\\
12-17, April, 2026\\
Bayreuth, Germany\\}

\begin{document}

\renewcommand{\hookAfterAbstract}{%
\par\bigskip
\textsc{SI-HEP-2026-20, P3H-26-068}
}

\maketitle

%%%%%%%%%%%%%%%%%%%%%%%%%%%%%%%%%%%%%%%%%%%%%%%%%%%%%%%%%%%%%%%%%%%%%%%
%%%%%%%%%%%%%%%%%%%%%%%%%%%%%%%%%%%%%%%%%%%%%%%%%%%%%%%%%%%%%%%%%%%%%%%

\section{Introduction}
\label{sec:intro}

In the quest for physics beyond the Standard Model (SM) decays of bottom-flavoured hadrons that proceed via flavour-changing neutral current (FCNC) processes play an important r\^ole since they probe scales far beyond the weak scale through virtual effects. In this context, the inclusive weak radiative decay $\bar B\to X_s \gamma$ of a $B$-meson into a charmless hadronic system $X_s$ of net strangeness $|S|=1$ and a photon has been established as one of the standard candles since high precision can be achieved both, on the experimental and theoretical side.

On the experimental side the current world average for the CP- and isospin-averaged branching ratio
${\mathcal B}_{s \gamma}$ reads~\cite{ParticleDataGroup:2024cfk,HeavyFlavorAveragingGroupHFLAV:2024ctg}
\be \label{eq:brexp} 
{\mathcal B}_{s \gamma}^{\rm exp} = (3.49 \pm 0.19)\times 10^{-4},
\ee
for a photon energy $E_\gamma > E_0 = 1.6\,{\rm GeV}$ in the rest-frame of the decaying $B$ meson. The current experimental uncertainty of $\pm 5.4\%$ is envisaged to be reduced to $\pm 2.6\%$ by the end of Belle~II~\cite{Belle-II:2018jsg,Ishikawa:2019TalkLyon}.

Also on the theory side tremendous effort and progress has been achieved since the next-to-leading order (NLO) prediction from 2001~\cite{Gambino:2001ew} which had an uncertainty of $\pm 8.3\%$. The major updates since then take next-to-next-to leading order (NNLO) effects into account and date from 2006~\cite{Misiak:2006zs}, 2015~\cite{Misiak:2015xwa}, and 2020~\cite{Misiak:2020vlo}. One of the main bottlenecks to reduce the overall uncertainty has been the charm-quark mass dependence of the interference between current-current and magnetic dipole operators at NNLO. In~\cite{Misiak:2006zs,Misiak:2006ab} the large-$m_c$ limit was calculated and extrapolated to physical values of~$m_c$. In~\cite{Misiak:2015xwa,Czakon:2015exa} the calculation was supplemented by the value at $m_c=0$ which turned the extrapolation into an interpolation. In~\cite{Misiak:2020vlo} the fermionic part of the interference was given for physical values of $m_c$. Still, a $\pm 3\%$ uncertainty has constantly been assigned to the $m_c$-extrapolation. In recent updates~\cite{Greub:2023msv,Czaja:2023ren,Fael:2023gau,Greub:2024mwp} this source of uncertainty got removed by calculating the interference in question for physical values of the charm-quark mass, and the new SM theory update was presented~\cite{Misiak:2026sqy}. In the present article we report on this calculation~\cite{Czaja:2026kop} and on multi-parton contributions at NLO~\cite{Brune:2025zhd}, a calculation that formally completes $\bar{B} \rightarrow X_s \gamma$ at NLO in QCD at leading power.
Both calculations are performed in the framework of the effective weak theory with
\begin{equation}
     \mathcal{L}_{\rm eff}=\mathcal{L}_{\rm QED+QCD}+\frac{4 G_F}{\sqrt{2}} V_{ts}^* V_{tb}
\left[ - \sum_{i=1}^2 C_i \sum_{p=u,c}\frac{ V_{ps}^* V_{pb}}{ V_{ts}^* V_{tb}} \, Q_i^p + \sum_{i=3}^8 C_{i} Q_i  \right] +\rm{h.c.} \, ,
\end{equation}
and with the corresponding effective operators defined as~\cite{Chetyrkin:1996vx}
\begin{align}
Q_1^p   &=  (\bar{s}_L \gamma_{\mu} T^a p_L) (\bar{p}_L \gamma^{\mu} T^a b_L), & Q_5  & =  (\bar{s}_L \gamma_{\mu}\gamma_{\nu}\gamma_{\sigma} b_L)\textstyle{\sum_q} (\bar{q} \gamma^{\mu}\gamma^{\nu}\gamma^{\sigma}     q),  \nonumber \\
Q_2^p   &=  (\bar{s}_L \gamma_{\mu}     p_L) (\bar{p}_L \gamma^{\mu}     b_L), & Q_6  & =  (\bar{s}_L \gamma_{\mu}\gamma_{\nu}\gamma_{\sigma} T^a b_L)\textstyle{\sum_q} (\bar{q} \gamma^{\mu}\gamma^{\nu} \gamma^{\sigma} T^a q), \nonumber \\
Q_3  & = (\bar{s}_L \gamma_{\mu}     b_L) \textstyle{\sum_q} (\bar{q}\gamma^{\mu} q), & Q_7  & =   \displaystyle{\frac{e}{16 \pi^2}} m_b (\bar{s}_L \sigma^{\mu \nu}     b_R) F_{\mu \nu} , \nonumber \\
Q_4  & =  (\bar{s}_L \gamma_{\mu} T^a b_L) \textstyle{\sum_q} (\bar{q}\gamma^{\mu} T^a q), & Q_8  & =   \displaystyle{\frac{g}{16 \pi^2}} m_b (\bar{s}_L \sigma^{\mu \nu} T^a b_R) G_{\mu \nu}^a.
\end{align}
The remainder of this article is organized as follows. In section~\ref{sec:multibody} we describe the calculation of multi-parton contributions at NLO, and in section~\ref{sec:exactmc} on the calculation of the $Q_{1,2}-Q_7$ interference at NNLO and physical value of $m_c$. We present the update of ${\cal B}_{s\gamma}$ in section~\ref{sec:newSMpredicton} and conclude in section~\ref{sec:conclusion}.

%%%%%%%%%%%%%%%%%%%%%%%%%%%%%%%%%%%%%%%%%%%%%%%%%%%%%%%%%%%%%%%%%%%%%%%
%%%%%%%%%%%%%%%%%%%%%%%%%%%%%%%%%%%%%%%%%%%%%%%%%%%%%%%%%%%%%%%%%%%%%%%

\section{Multi-body contributions to NLO}
\label{sec:multibody}

\begin{figure}
  \begin{center}
	\includegraphics[width=0.32\textwidth]{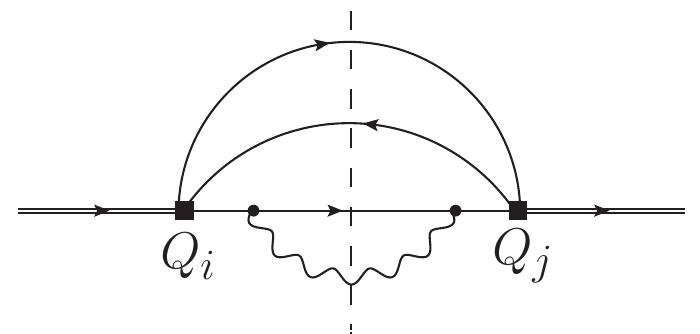}
	\includegraphics[width=0.32\textwidth]{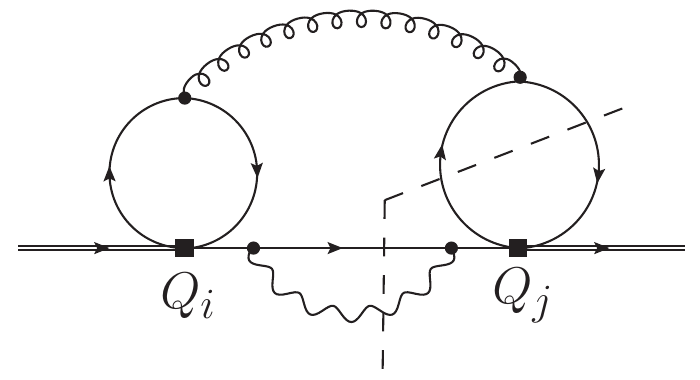}
	\includegraphics[width=0.32\textwidth]{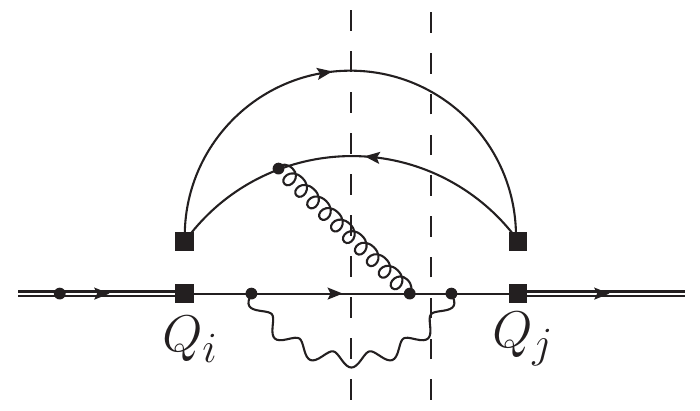}
  \end{center}
  \caption{Left: Tree-level contributions to $b \rightarrow s q\bar{q}\gamma$~\cite{Kaminski:2012eb}. Middle: Pure four-body $b \to s  q  \bar q  \gamma$ contributions at NLO~\cite{Huber:2014nna}. Right: Four- and five-body $b \rightarrow s q\bar{q}(g)\gamma$ contributions at NLO~\cite{Brune:2025zhd}. The black squares denote operator insertions $Q_{1,\ldots,6}$ from the effective Hamiltonian.\label{fig:5bdiags}}
\end{figure}

In the framework of the Heavy-Quark Expansion, the decay width of the inclusive decay $\bar{B} \rightarrow X_s \gamma$ with a photon-energy cut $E_0$ can be written as
\begin{equation}
    \Gamma(\bar{B} \rightarrow X_s \gamma)_{E_{\gamma}>E_0}=\Gamma(b \rightarrow X_s^{\rm parton} \gamma)_{E_{\gamma}>E_0}+\mathcal{O}(\Lambda_{\rm{QCD}}/m_b) \, .
\end{equation}
The first term on the right-hand side can be further decomposed as
\begin{eqnarray}
    \Gamma(b \rightarrow X_s^{\rm parton} \gamma)_{E_{\gamma}>E_0}& = &\Gamma(b \rightarrow s \gamma)+\Gamma(b \rightarrow s g \gamma)+\Gamma(b \rightarrow s q\bar{q}\gamma)+\Gamma(b \rightarrow s q\bar{q}g\gamma)+\ldots \nonumber \\[0.4em]
    &=& \frac{G_F^2 m_b^5 \alpha_e |V_{ts}^{*}V_{tb}|^2}{32 \pi^4} \sum_{i,j} \mathcal{C}_i^{\text{eff} \, *}(\mu) \, \mathcal{C}_j^{\text{eff}}(\mu) \,  \hat{G}_{ij}(\mu,z_c,\delta) \, .
\end{eqnarray}
The $\mathcal{C}_i^{\text{eff}}$ are linear combinations of Wilson coefficients, $z_c=m_c^2/m_b^2$ denotes the dependence on the charm-quark mass, and the photon-energy cut will be parametrized by the dimensionless variable $\delta = 1-2E_0/m_b$ in the following. Here, we focus on the indicated four- and five-body contributions $b \rightarrow s q\bar{q}(g)\gamma$, where $q \in \{u,d,s\}$ is a light quark. The tree-level contributions to $b \rightarrow s q\bar{q}\gamma$ were computed in~\cite{Kaminski:2012eb}, while those NLO contributions that require four-body $b \to s  q \bar q \gamma$ final states only were obtained in~\cite{Huber:2014nna}. In a recent work~\cite{Brune:2025zhd}, we computed those one-loop four-particle $b \to s q \bar{q} \gamma$ diagrams that must be supplemented by the corresponding five-particle tree-level $b \to s q \bar{q}  g  \gamma $ cuts originating from gluon bremsstrahlung. The three types of contributions are displayed in figure~\ref{fig:5bdiags}. Ref.~\cite{Brune:2025zhd} represents the last piece that was missing in order to formally complete $\bar{B} \rightarrow X_s \gamma$ at NLO in QCD at leading power.

\begin{figure}
  \begin{center}
    \includegraphics[width=0.46\textwidth]{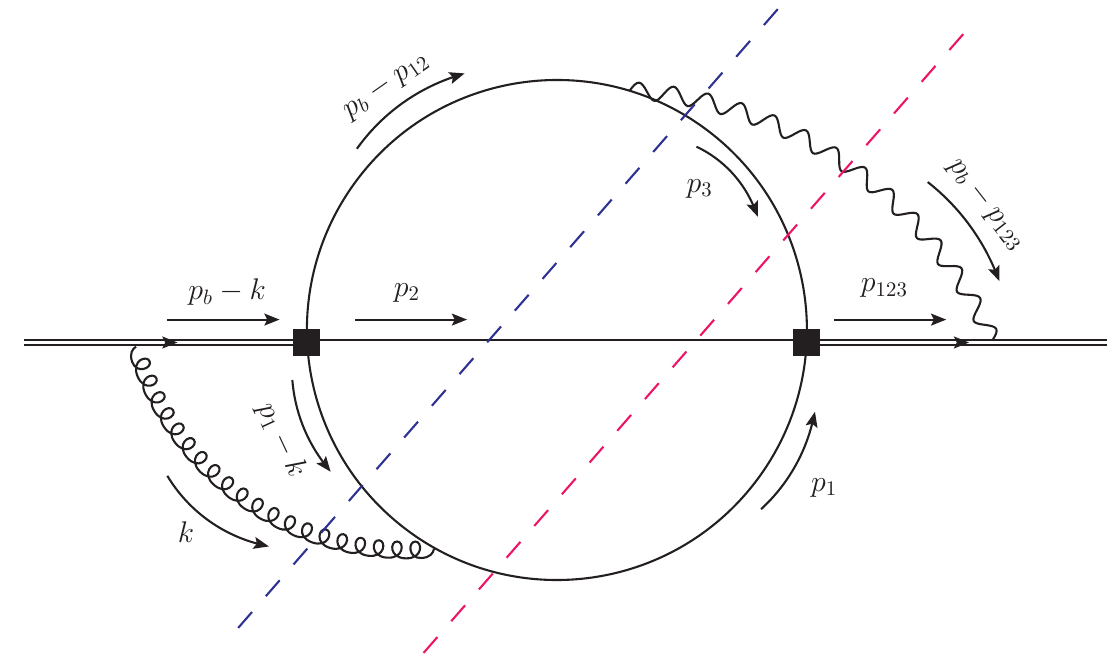} \qquad
    \includegraphics[width=0.48\textwidth]{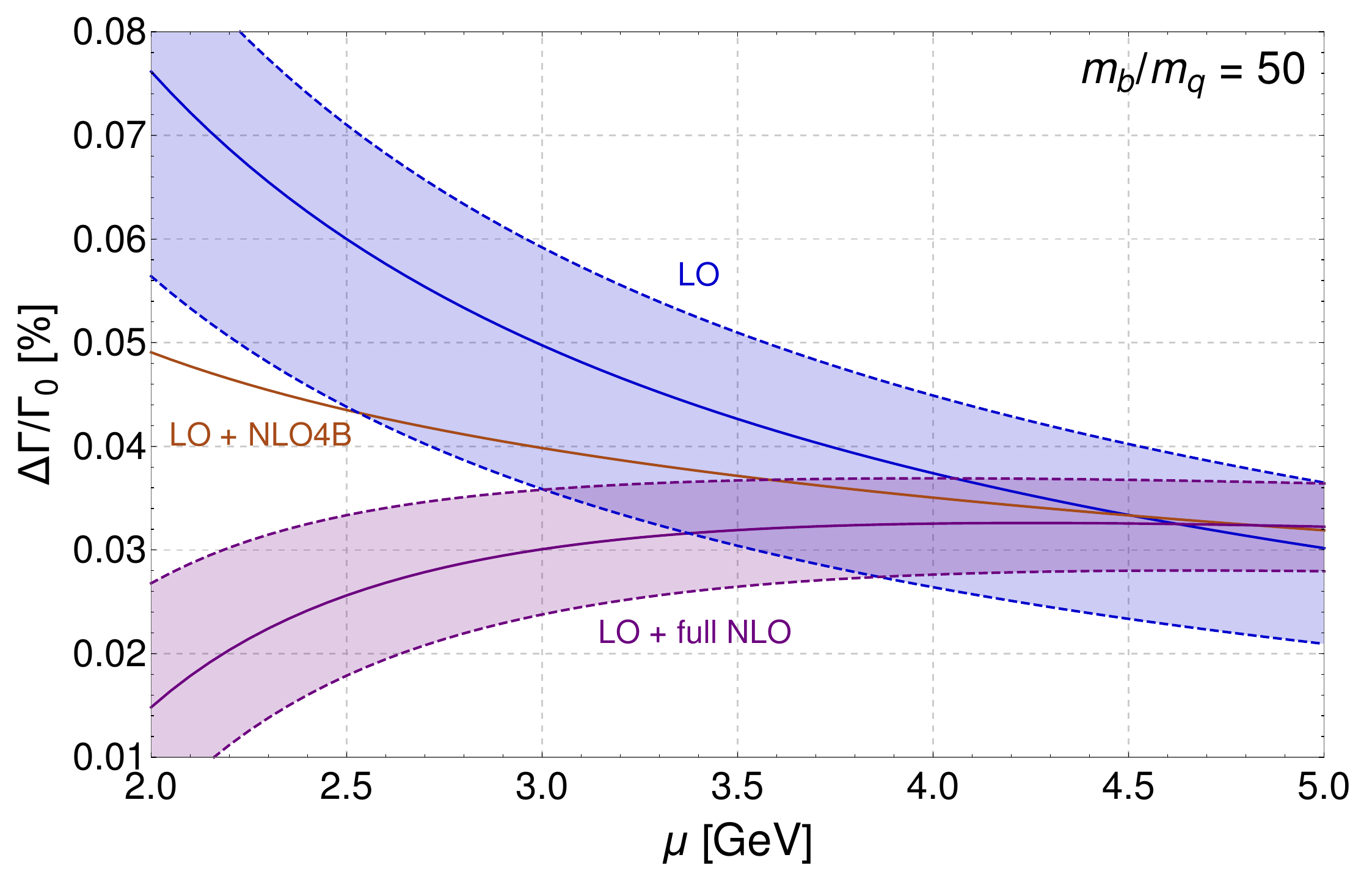}%plotE050.pdf
  \end{center}
  \caption{Left: A sample diagram which fits into topology ${\cal{T}}_2$, together with a four (red) and five (blue) particle cut. Right: Renormalisation-scale dependence of the normalised decay rate $\Delta\Gamma/\Gamma_0$ at leading order (LO, blue), next-to-leading order including only the four-body contributions (LO~+~NLO4B, brown), and including in addition the contributions calculated in~\cite{Brune:2025zhd} (LO~+~full~NLO). \label{fig:5bnlofull}}
\end{figure}

\begin{figure}
  \begin{center}
    \includegraphics[width=0.43\textwidth]{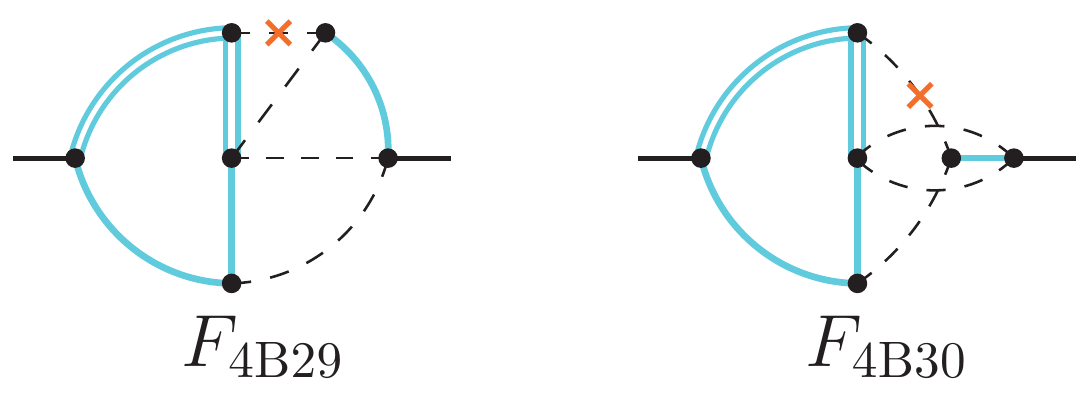} \qquad
    \includegraphics[width=0.44\textwidth]{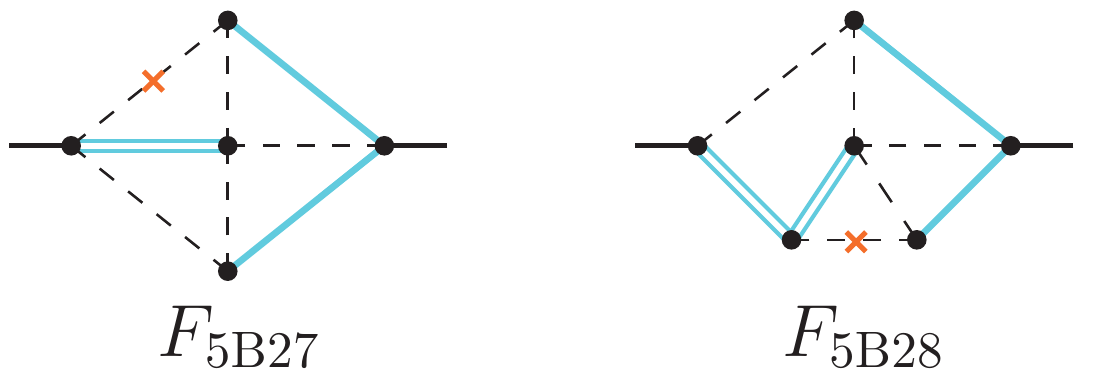}%plotE050.pdf
  \end{center}
  \caption{Sample four- and five-body master integrals. Dashed lines indicate cut propagators, the solid light blue single (double) lines indicate massless (massive) propagators. The orange cross denotes the cut photon propagator. Due to the energy cut, this line breaks the symmetry in the final-state momenta. \label{fig:masters}}
\end{figure}

The actual calculation in ref.~\cite{Brune:2025zhd} amounts to the computation of $\sim$$180$ one-loop four-body $b \rightarrow s q\bar{q}\gamma$ diagrams and $\sim$$400$ tree-level five-body $b \to s q  \bar{q}  g  \gamma $ diagrams per $Q_i-Q_j$ interference. In order to incorporate the cut on the photon energy we stay differential in the variable
\begin{equation}
\frac{2 \, E_\gamma}{m_b} = \frac{2 \, p_b \cdot p_\gamma}{m_b^2} \equiv 1-z \equiv \bar z \, ,
\end{equation}
which we implement at the integrand level via a factor $\displaystyle \delta(\bar z - 2 p_b \cdot p_\gamma/m_b^2)$. We then turn this $\delta$-function and the ones originating from cut lines in the phase-space integration into loop-integral propagators via reverse unitarity~\cite{Anastasiou:2002yz},
\begin{equation}
-2 \pi i \, \delta(A) = \frac{1}{A+i\eta} - \frac{1}{A-i\eta} \, .
\end{equation}
This procedure converts all integrals into partially massive four-loop propagator-type integrals. They can be cast into seven integral topologies ${\cal{T}}_{1,\ldots,7}$, e.g.
\begin{align}
{\cal{T}}_2 & = \Big\{ (p_b - p_{123})^2, 2p_b(p_b-p_{123})-m_b^2 \, \bar z,p_1^2,p_2^2,p_3^2,k^2,(p_3-k)^2,(p_{123}-k)^2-m_b^2, \nonumber \\[0.6em]
& \qquad (p_b-k)^2-m_b^2, (p_b-p_{12})^2,(p_1-k)^2,(p_b-p_{13})^2,(p_b-p_1)^2,(p_b-p_3)^2\Big\}
\end{align}
which has three massive propagators and two different masses in total. A sample diagram which fits into this topology is displayed in the left panel of figure~\ref{fig:5bnlofull}. The integrals then undergo an integration-by-parts (IBP) reduction~\cite{Chetyrkin:1981qh,Tkachov:1981wb,Laporta:2000dsw} using the program \texttt{FIRE}~\cite{Smirnov:2019qkx}, which results in $\sim$$30$ four-body master integrals and about the same number of five-particle ones. A few representatives are shown in figure~\ref{fig:masters}.

For the actual calculation of the master integrals we apply several techniques. The easier ones are directly integrated over the four- respectively five-particle massless phase-space in $D=4-2\epsilon$ dimensions~\cite{Gehrmann-DeRidder:2003pne,Heinrich:2006sw}. For all other integrals we employ the method of differential equations~\cite{Kotikov:1990kg,Remiddi:1997ny,Argeri:2007up}, partially in a canonical basis~\cite{Henn:2013pwa} to which we convert by means of the package \texttt{epsilon}~\cite{Prausa:2017ltv}. The boundary conditions are obtained from asymptotic expansions as $z \to 0$ or~$1$, which result in hypergeometric functions or Mellin-Barnes representations which can subsequently be expanded in $\epsilon$~\cite{Huber:2005yg,Czakon:2005rk}. Another technique to obtain the boundary condition is to remove the photon-energy cut by integrating over the entire phase space~\cite{Gituliar:2015iyq}, i.e.\ we don't stay differential in $\bar z$ but integrate over $z = 0\ldots 1$. Taking all these techniques together, we can solve all master integrals analytically in terms of harmonic and Goncharov polylogarithms. The alphabet of the iterated integrals reads $\{0,\pm 1\, \pm i/\sqrt{3}\}$, and the arguments of the generalized polylogarithms are from the set $\{\bar z,i\sqrt\frac{\bar z}{4-\bar z}\}$.

After these steps the decay width is still ultraviolet (UV) and infrared (IR) divergent. The UV divergences are cancelled via standard renormalisation at order ${\cal O}(\alpha_s)$, which involves mass, wave-function and operator renormalisation. The latter also needs to take into account mixing between operators and inclusion of evanescent operators to make the system closed under renormalisation. The IR divergences are of collinear nature and stem from the region of phase space where the photon becomes collinear to a massless quark. Here we apply the same method as in refs.~\cite{Huber:2005ig,Huber:2014nna,Kaminski:2012eb}, where a universal $q \to q\gamma$ splitting kernel is convoluted with the $b \to s q \bar q (g)$ decay width which lacks the photon. Since the splitting kernel contains both, the dimensional regulator and the mass of the quark, 
this procedure allows to translate the collinear $1/\epsilon$ pole into a logarithm $\log(m_b^2/m_q^2)$ of the light quark masses via the prescription
\begin{equation}
    \label{eq:DecayShiftprescription}
    \frac{\mathrm{d}\Gamma_{m}}{\mathrm{d}z}=\frac{\mathrm{d}\Gamma_{\eps}}{\mathrm{d}z}+\frac{\mathrm{d}\Gamma_{\text{shift}}}{\mathrm{d}z}\, .
\end{equation}
Note that in our NLO calculation we need the convolution of the LO splitting kernel with the NLO $b \to s q \bar q (g)$ decay width, and that of the NLO splitting kernel with the LO $b \to s q \bar q$ width. All the required kernels are given explicitly in ref.~\cite{Brune:2025zhd}. After this procedure all poles in $\epsilon$ cancel and we obtain a finite expression for the decay width differential in $\bar z$ which we subsequently integrate over $\bar z = \bar\delta \ldots 1$. We also checked that the result is renormalisation-group invariant to the order we are working.

To investigate the size of the correction we evaluate the multibody contribution $\Delta\Gamma$ relative to the leading-order two-body $b\to s \gamma$ decay width, $\Gamma_0 |\mathcal{C}_{7}^{\text{(0)eff}}|^2$, where
\begin{equation}
\label{eq:normalisationfactor}
\Gamma_0 = \frac{G_F^2 m_b^5 \alpha_e |V_{ts}^{*}V_{tb}|^2}{32 \pi^4}
\end{equation}
and $|\mathcal{C}_{7}^{\text{(0)eff}}(\mu_b)|^2 \sim 0.1$. Depending on the ratio $m_b/m_q$ we obtain numbers in the range of a percent or slightly below, e.g.
\begin{align}
\Delta\Gamma/\Gamma_{0}|\mathcal{C}_{7}^{\text{(0)eff}}|^2{}_{\big|_{m_q=m_b/50}} & = \, 0.1899~\% \, .
\end{align}
The smallness of the result was partially expected due to CKM and phase-space suppression and small Wilson coefficients. In addition, both, the contributions calculated in~\cite{Huber:2014nna} and~\cite{Brune:2025zhd} are negative and thus lower the value found at LO, see right panel in figure~\ref{fig:5bnlofull}. As can be also seen from that plot, the scale dependence and total uncertainty is reduced at NLO compared to LO, as expected. The absolute impact of the multi-body contribution on ${\cal{B}}_{s\gamma}$ will be given in section~\ref{sec:newSMpredicton}.

%%%%%%%%%%%%%%%%%%%%%%%%%%%%%%%%%%%%%%%%%%%%%%%%%%%%%%%%%%%%%%%%%%%%%%%
%%%%%%%%%%%%%%%%%%%%%%%%%%%%%%%%%%%%%%%%%%%%%%%%%%%%%%%%%%%%%%%%%%%%%%%

\section{\boldmath Exact $m_c$-dependence of the $Q_{1,2}-Q_7$ interference at NNLO}
\label{sec:exactmc}

\begin{figure}
  \begin{center}
    \includegraphics[width=0.98\textwidth]{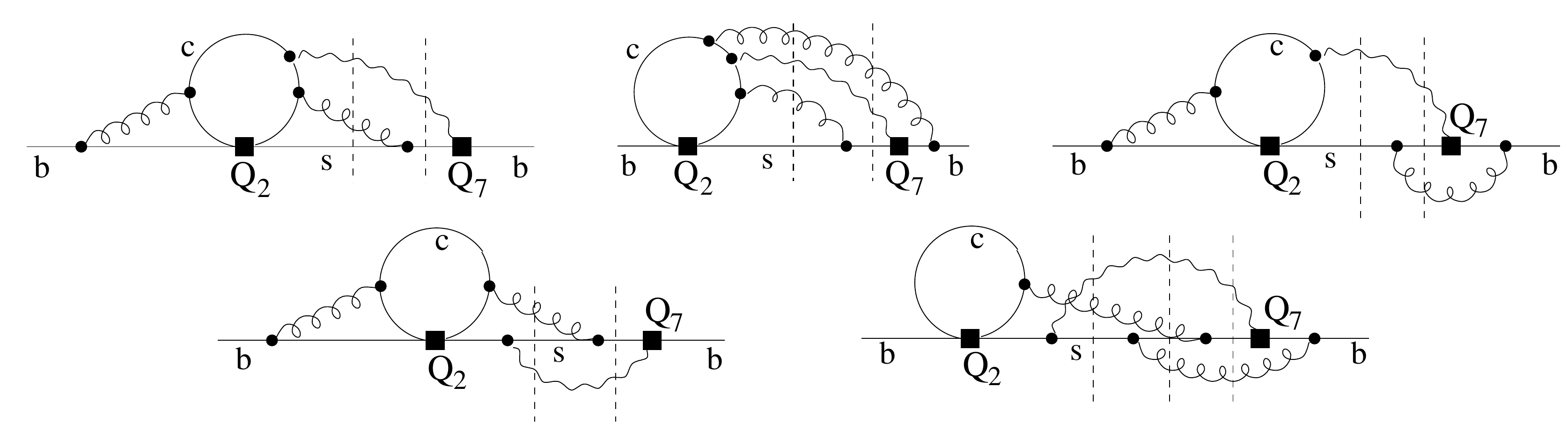}
  \end{center}
  \caption{Sample diagrams of the $Q_{2}-Q_7$ interference at NNLO. The vertical dashed lines denote unitarity cuts. \label{fig:diagsbsg}}
\end{figure}

As mentioned above, to remove the $\pm 3\%$ uncertainty assigned to the interpolation in $m_c$, the $Q_{1,2}-Q_7$ interference at NNLO has to be computed for physical values of the charm-quark mass~\cite{Czaja:2026kop}. To this end several hundred four-loop propagator diagrams with two-, three- and four-particle cuts have to be computed, a sample of which is depticted in figure~\ref{fig:diagsbsg}. After Dirac- and colour-algebra they result in several hundred-thousand four-loop, two-scale scalar integrals with unitarity cuts in ${\cal O}(500)$ families. They undergo an integral reduction using Kira~2.0~\cite{Klappert:2020nbg} and FIRE~6~\cite{Smirnov:2019qkx}, which takes up to a few weeks of CPU time and ${\cal O}(1\textrm{TB})$ of RAM per topology. The reduction results in ${\cal O}(500)$ master integrals, which we solve separately for the two-, three- and four-particle cuts by means of several variants of the method of differential equations (DE) w.r.t.\ the variable $z=m_c^2/m_b^2$.
The first method starts by calculating boundary conditions at $z = \infty$. These single-scale integrals are solved analytically by using hypergeometric functions~\cite{Huber:2005yg,Huber:2007dx}, Mellin-Barnes techniques~\cite{Czakon:2005rk} and the {\tt PSLQ}-algorithm~\cite{Ferguson:1999pslq}. Subsequently, deep expansions about $z = \infty$ are used as starting point for the numerical evolution along a contour in the complex plane. Going away from the real axis is required to circumvent the threshold at $z_{{\textrm{thr.}}}=1/4$. The actual numerical evolution is done with the library Odeint~\cite{Ahnert:2011ode} and the Fortran collection ODEPACK of ordinary DE solvers~\cite{Hindmarsh:1983ode}. A dedicated description of the method can be found in~\cite{Niggetiedt:974075}.
The second method applies the technique of ``expand and match'' developed in~\cite{Fael:2021kyg}. It uses the DEs to construct generic
expansions about properly chosen values of $z$ which are matched onto corresponding expansions about neighboring points inside the radius of convergence. The boundary conditions are typically obtained from {\tt AMFlow}~\cite{Liu:2022chg}. Finally, as a third method (\cite{Czaja:2023ren}, see also~\cite{Greub:2023msv,Fael:2023gau,Greub:2024mwp}) which can be applied to the two- and four-particle cuts since they are inert to the $i \eta$ prescription in the propagators, we can obtain the values for the master integrals in the physical region of $z$ directly with {\tt AMFlow}. The results of the various methods agree with each other, the final numerical accuracy for physical values of $z$ is about 10 digits.
\begin{figure}
  \begin{center}
    \includegraphics[width=0.48\textwidth]{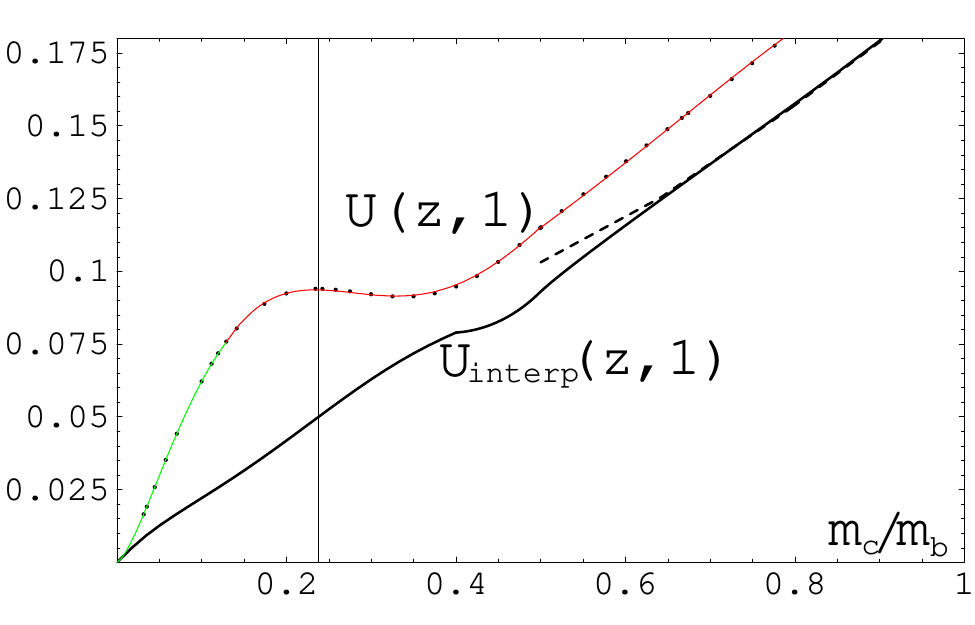} \qquad
    \includegraphics[width=0.46\textwidth]{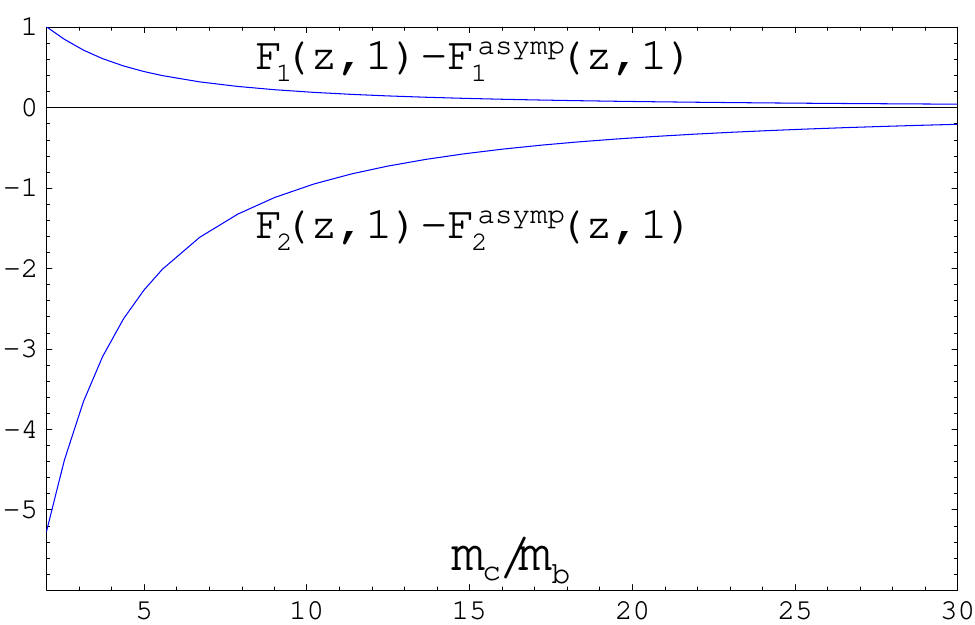}
  \end{center}
  \caption{Left: The interpolated (black) and exact (coloured) versions of the function $U(z,1)$. Right: The differences $F_{1,2}(z,1) - F_{1,2}^{\textrm{asymp}}(z,1)$ of the functions $F_{1,2}(z,1)$ in eq.~(\ref{eq:defU}). See text for further explanations. \label{fig:UF}}
\end{figure}

After renormalisation and the cancellation of dimensionally regulated divergences, the $Q_{1,2}-Q_7$ interference at NNLO can be described by a single function $U(z,\delta)$ defined as
\begin{equation}
\frac{\Delta{\mathcal B_{s\gamma}}}{\mathcal B_{s\gamma}^{\rm LO}} ~\simeq~ U(z,\delta) ~\equiv~
\frac{\alpha_s^2(\mu_b)}{8\pi^2}\;\;
\frac{ C_1^{(0)} F_1(z,\delta) + ( C_2^{(0)} -\frac{1}{6}C_1^{(0)} ) F_2(z,\delta)}{
    C_7^{(0)}-\frac{1}{3} C_3^{(0)}-\frac{4}{9} C_4^{(0)}-\frac{20}{3} C_5^{(0)}-\frac{80}{9} C_6^{(0)}} \, . \label{eq:defU}
\end{equation}
It is shown in the left panel of figure~\ref{fig:UF} and draws in black the interpolated function used so far, and in colour the newly-obtained exact function, which exhibits several interesting features. First, the point $m_c=0$ from~\cite{Czakon:2015exa} is reproduced. Second the difference between the exact and the interpolated function is largest around the physical value of $z$, indicated by the vertical line. This will also have a sizeable impact on the central value of ${\cal B}_{s\gamma}$, as we will quantify in the next section. Finally, we remark that the convergence towards the asymptotic curve for large $z$ from~\cite{Misiak:2006ab} is slow, as can be seen from the right panel in figure~\ref{fig:UF} which shows the difference $F_{1,2}(z,1) - F_{1,2}^{\textrm{asymp}}(z,1)$ for the individual functions $F_{1,2}(z,1)$ from eq.~(\ref{eq:defU}). The slowest-converging term for each difference turns out to be proportional to $\log^2(z)/z$.

%%%%%%%%%%%%%%%%%%%%%%%%%%%%%%%%%%%%%%%%%%%%%%%%%%%%%%%%%%%%%%%%%%%%%%%
%%%%%%%%%%%%%%%%%%%%%%%%%%%%%%%%%%%%%%%%%%%%%%%%%%%%%%%%%%%%%%%%%%%%%%%

\section{\boldmath Update of the Standard-Model prediction for ${\cal{B}}_{s\gamma}$}
\label{sec:newSMpredicton}

The transition from the previous Standard-Model prediction for ${\cal{B}}_{s\gamma}$~\cite{Misiak:2015xwa} to the new one proceeds in several steps, which we outline in the following. For each of them we quantify the respective change in table~\ref{tab:brshifts}.
\begin{enumerate}
\item We change the normalisation from
\begin{equation}
{\cal B}(\bar B \to X_s\gamma)_{E_\gamma>E_0}  = {\cal B}(\bar B \to X_c \ell \bar \nu)_{{\textrm{exp}}} \, \left|\frac{V_{ts}^\ast V_{tb}}{V_{cb}}\right|^2 \, \frac{6\alpha_{{\textrm{em}}}}{\pi \, C} \big[ P(E_0) + N(E_0)\big]
\end{equation}
to
\begin{equation}
{\cal B}_{s\gamma} = \frac{G_F^2 \, \alpha_{{\textrm{em}}} \, {m^5_{b,{\textrm{kin}}}} \, \tau_{{\textrm{av}}}}{32\pi^4} \left|V_{ts}^\ast V_{tb}\right|^2 \, \big[ \widetilde P(E_0) + \widetilde N(E_0)\big]
\end{equation}
due to the poor behaviour of the perturbative series for the semi-leptonic phase-space factor
\begin{equation}
C = \left|\frac{V_{ub}}{V_{cb}}\right|^2 \, \frac{{\Gamma}(\bar B \to X_c \ell \bar \nu)}{{\Gamma}(\bar B \to X_u \ell \bar \nu)}
\end{equation}
at ${\cal O}(\alpha_s^3)$~\cite{Fael:2020tow,Chen:2023osm,Chen:2023dsi,Fael:2023tcv} and the precise knowledge of the kinetic mass $m_{b,{\textrm{kin}}}$~\cite{Fael:2020njb}.
\item We update all input parameters to their current values, see Appendix C of~\cite{Czaja:2026kop}. %\\[-1.6em]
\item We modify the treatment of non-perturbative resolved photon contributions following ref.~\cite{Misiak:2020vlo}, which extracts the value for $\kappa_V$ from ranges of~$\Lambda_{17}$ given in~\cite{Gunawardana:2019gep}. There are further recent developments on the non-perturbative resolved photon contributions~\cite{Benzke:2020htm,Bartocci:2024bbf,Benzke:2025ekp} which include partial ${\cal O}(\Lambda_{\textrm{QCD}}^2/m_b^2)$ corrections. Here, we prefer to stay at the ${\cal O}(\Lambda_{\textrm{QCD}}/m_b)$ level until the ${\cal O}(\Lambda_{\textrm{QCD}}^2/m_b^2)$ corrections get estimated in a complete manner.
\item We add the four- and five-body NLO contributions from ref.~\cite{Brune:2025zhd}.
\item Finally, we replace the previously interpolated NNLO functions $F_{1,2}(z,1)$ (respectively $U_{\textrm{interp.}}(z,1)$) by their precise behaviour in the physical region of $m_c$.
\end{enumerate}
\begin{table}[t]
\begin{center}
\begin{tabular}{|ccccc|c|}\hline
1 & 2 & 3 & 4 & 5 & total \\\hline
  $+1.5\%\!$ 
& $-1.1\%\!$ 
& $+1.1\%\!$ 
& $-0.2\%\!$ 
& $+4.2\%\!$ 
& $+5.5\%\!$ 
\\\hline
\end{tabular}
\end{center}
\ \\[-1cm]
\caption{Shifts in the central value of ${\mathcal B}_{s\gamma}$
for $E_0 = 1.6\,$GeV at each step (see the text).\label{tab:brshifts}}
\end{table}

\begin{figure}
  \begin{center}
    \includegraphics[width=0.51\textwidth]{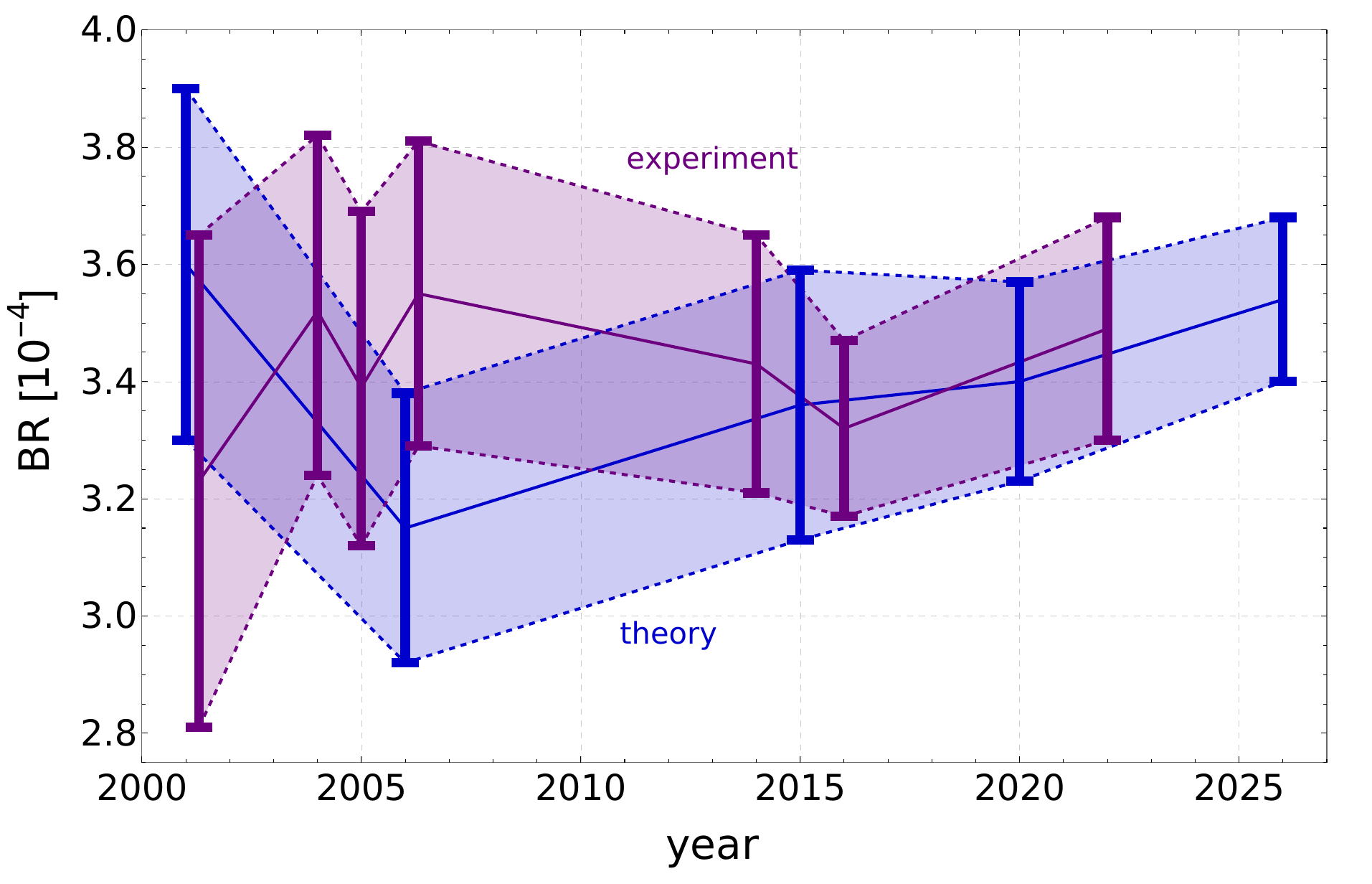}
  \end{center}
  \caption{Development of the SM prediction and experimental value of the branching ratio of $\bar B \to X_s\gamma$ over time. \label{fig:history2hdm}}
\end{figure}

As we anticipated before, the last change has the largest effect and is responsible for about three quarters of the total shift. Putting everything together, our final prediction for ${\mathcal B}_{s\gamma}$ in the SM reads~\cite{Czaja:2026kop,Misiak:2026sqy}
\be \label{eq:brsm}
{\mathcal B}_{s\gamma}^{\rm SM} = \left( 3.54 \pm 0.14 \right) \times 10^{-4}
\ee
for $E_\gamma > 1.6\,{\rm GeV}$. The total uncertainty of $\pm
4\%$ has been obtained by adding the parametric ($\pm 2.7\%$) and the
higher-order ($\pm 3 \%$) ones in quadrature. Note that the scale dependence is much smaller than $\pm 3 \%$. Yet, we keep that value to account for potentially sizeable but yet unknown higher-order effects in the power-suppressed, non-perturbative corrections.
The result agrees very well with the experimental value in eq.~(\ref{eq:brexp}).
It is remarkable that the central value of the new NNLO result in eq.~(\ref{eq:brsm})
is only $1.7\%$ lower than the NLO one presented a quarter of a century ago in
ref.~\cite{Gambino:2001ew}, yet with an uncertainty that has been cut to half by now.
The fact that intermediate predictions had significantly smaller central values
reflects the rough character of the interpolation, as discussed before. The time-evolution of the SM prediction and experimental value is graphically summarized in the left panel of figure~\ref{fig:history2hdm}.

The SM prediction, together with the experimental value, can be used to put a constraint on the mass $M_{H^\pm}$ of the charged Higgs-boson in type-II Two-Higgs-Doublet models. Using the numbers in eqs.~(\ref{eq:brexp}) and~(\ref{eq:brsm}) results in
\be \label{eq:MH2hdm}
 M_{H^\pm} > 670~{\textrm{GeV}} \quad {\textrm{at}} \quad 95\%~{\textrm{C.L.}} \, .
\ee

%%%%%%%%%%%%%%%%%%%%%%%%%%%%%%%%%%%%%%%%%%%%%%%%%%%%%%%%%%%%%%%%%%%%%%%
%%%%%%%%%%%%%%%%%%%%%%%%%%%%%%%%%%%%%%%%%%%%%%%%%%%%%%%%%%%%%%%%%%%%%%%

\section{Conclusion and outlook}
\label{sec:conclusion}

The inclusive radiative FCNC decay $\bar B \to X_s \gamma$ will remain an important player in the field of precision quark flavour physics and in beyond-SM studies. In the present article we report on two recent calculations, multi-parton contributions at NLO~\cite{Brune:2025zhd} and the calculation of the $Q_{1,2}-Q_7$ interference at NNLO and physical value of $m_c$~\cite{Czaja:2026kop}. The former calculation formally completes $\bar{B} \rightarrow X_s \gamma$ at NLO in QCD at leading power, the latter removes the $\pm 3\%$ uncertainty arising from interpolation in $m_c$ and is responsible for a sizeable shift in the central value, as can be seen from the update of ${\cal B}_{s\gamma}$ in eq.~(\ref{eq:brsm}) and the numbers in table~\ref{tab:brshifts}. The result in eq.~(\ref{eq:brsm}) has a total uncertainty of $\pm 4\%$ which combines the parametric ($\pm 2.7\%$) and the higher-order ($\pm 3 \%$) ones in quadrature.

As interesting future directions, the $m_c$ dependence of the $Q_{3-6}-Q_{7}$ interferences can be calculated at NNLO. Moreover, up to now the NNLO $Q_{1-6}-Q_{7}$ interferences have been calculated without a cut on the photon energy, which would represent an additional scale in the problem. With contemporary tools such a calculation is certainly feasible. Furthermore, first steps towards ${\cal O}(\alpha_s^3)$ have been undertaken~\cite{Fael:2026fxp} and will be pursued further in the future. Another prominent r\^ole will be played by nonperturbative resolved-photon corrections in the $Q_{1,2}-Q_{7}$ and $Q_{8}-Q_{8}$ interference~\cite{Hurth:2023paz,Gunawardana:2019gep,Benzke:2020htm,Bartocci:2024bbf,Benzke:2025ekp}.

%%%%%%%%%%%%%%%%%%%%%%%%%%%%%%%%%%%%%%%%%%%%%%%%%%%%%%%%%%%%%%%%%%%%%%%
%%%%%%%%%%%%%%%%%%%%%%%%%%%%%%%%%%%%%%%%%%%%%%%%%%%%%%%%%%%%%%%%%%%%%%%

\section*{Acknowledgments}

I would like to warmly thank my co-authors from refs.~\cite{Brune:2025zhd,Czaja:2026kop} for their collaboration, and the organisers of LL2026 for creating a very pleasant and inspiring atmosphere. This research was supported by Deutsche Forschungsgemeinschaft (DFG, German Research Foundation) under grant 396021762 --- TRR 257 ``Particle Physics Phenomenology after the Higgs Discovery''. Support from Deutsche Forschungsgemeinschaft (DFG, German Research Foundation) under Germany's Excellence Strategy~--~Cluster of Excellence ``Color meets Flavor'', EXC 3107~--~Project-ID 533766364 is also acknowledged.

%%%%%%%%%%%%%%%%%%%%%%%%%%%%%%%%%%%%%%%%%%%%%%%%%%%%%%%%%%%%%%%%%%%%%%%
%%%%%%%%%%%%%%%%%%%%%%%%%%%%%%%%%%%%%%%%%%%%%%%%%%%%%%%%%%%%%%%%%%%%%%%

%\begin{thebibliography}{99}
%\bibitem{...}

\bibliographystyle{JHEP}
\bibliography{references}

\providecommand{\href}[2]{#2}\begingroup\raggedright\begin{thebibliography}{10}

\bibitem{ParticleDataGroup:2024cfk}
{\scshape Particle Data Group} collaboration, \emph{{Review of particle
  physics}}, \href{https://doi.org/10.1103/PhysRevD.110.030001}{\emph{Phys.
  Rev. D} {\bfseries 110} (2024) 030001}.

\bibitem{HeavyFlavorAveragingGroupHFLAV:2024ctg}
{\scshape Heavy Flavor Averaging Group (HFLAV)} collaboration, \emph{{Averages
  of b-hadron, c-hadron, and {\ensuremath{\tau}}-lepton properties as of
  2023}}, \href{https://doi.org/10.1103/x87q-tld5}{\emph{Phys. Rev. D}
  {\bfseries 113} (2026) 012008}
  [\href{https://arxiv.org/abs/2411.18639}{{\ttfamily 2411.18639}}].

\bibitem{Belle-II:2018jsg}
{\scshape Belle-II} collaboration, \emph{{The Belle II Physics Book}},
  \href{https://doi.org/10.1093/ptep/ptz106}{\emph{PTEP} {\bfseries 2019}
  (2019) 123C01} [\href{https://arxiv.org/abs/1808.10567}{{\ttfamily
  1808.10567}}].

\bibitem{Ishikawa:2019TalkLyon}
A.~Ishikawa. {Talk at the ``7th Workshop on Rare Semileptonic $B$ Decays'',
  September 4-6th, 2019, Lyon,
  France.~}\url{https://indico.in2p3.fr/event/18646}.

\bibitem{Gambino:2001ew}
P.~Gambino and M.~Misiak, \emph{{Quark mass effects in anti-B
  ---{\ensuremath{>}} X(s gamma)}},
  \href{https://doi.org/10.1016/S0550-3213(01)00347-9}{\emph{Nucl. Phys. B}
  {\bfseries 611} (2001) 338}
  [\href{https://arxiv.org/abs/hep-ph/0104034}{{\ttfamily hep-ph/0104034}}].

\bibitem{Misiak:2006zs}
M.~Misiak et~al., \emph{{Estimate of $\mathcal{B} (\bar B \to X_s \gamma)$ at
  $O(\alpha_s^2)$}},
  \href{https://doi.org/10.1103/PhysRevLett.98.022002}{\emph{Phys. Rev. Lett.}
  {\bfseries 98} (2007) 022002}
  [\href{https://arxiv.org/abs/hep-ph/0609232}{{\ttfamily hep-ph/0609232}}].

\bibitem{Misiak:2015xwa}
M.~Misiak et~al., \emph{{Updated NNLO QCD predictions for the weak radiative
  B-meson decays}},
  \href{https://doi.org/10.1103/PhysRevLett.114.221801}{\emph{Phys. Rev. Lett.}
  {\bfseries 114} (2015) 221801}
  [\href{https://arxiv.org/abs/1503.01789}{{\ttfamily 1503.01789}}].

\bibitem{Misiak:2020vlo}
M.~Misiak, A.~Rehman and M.~Steinhauser, \emph{{Towards $ \overline{B}\to
  {X}_s\gamma $ at the NNLO in QCD without interpolation in m$_{c}$}},
  \href{https://doi.org/10.1007/JHEP06(2020)175}{\emph{JHEP} {\bfseries 06}
  (2020) 175} [\href{https://arxiv.org/abs/2002.01548}{{\ttfamily
  2002.01548}}].

\bibitem{Misiak:2006ab}
M.~Misiak and M.~Steinhauser, \emph{{NNLO QCD corrections to the $\bar B \to
  X_s \gamma$ matrix elements using interpolation in $m_c$}},
  \href{https://doi.org/10.1016/j.nuclphysb.2006.11.027}{\emph{Nucl. Phys. B}
  {\bfseries 764} (2007) 62}
  [\href{https://arxiv.org/abs/hep-ph/0609241}{{\ttfamily hep-ph/0609241}}].

\bibitem{Czakon:2015exa}
M.~Czakon, P.~Fiedler, T.~Huber, M.~Misiak, T.~Schutzmeier and M.~Steinhauser,
  \emph{{The $(Q_{7}, Q_{1,2})$ contribution to $ \overline{B}\to {X}_s\gamma $
  at $ \mathcal{O}\left({\alpha}_{\mathrm{s}}^2\right) $}},
  \href{https://doi.org/10.1007/JHEP04(2015)168}{\emph{JHEP} {\bfseries 04}
  (2015) 168} [\href{https://arxiv.org/abs/1503.01791}{{\ttfamily
  1503.01791}}].

\bibitem{Greub:2023msv}
C.~Greub, H.M.~Asatrian, F.~Saturnino and C.~Wiegand, \emph{{Specific
  three-loop contributions to b {\textrightarrow} s{\ensuremath{\gamma}}
  associated with the current-current operators}},
  \href{https://doi.org/10.1007/JHEP05(2023)201}{\emph{JHEP} {\bfseries 05}
  (2023) 201} [\href{https://arxiv.org/abs/2303.01714}{{\ttfamily
  2303.01714}}].

\bibitem{Czaja:2023ren}
M.~Czaja, M.~Czakon, T.~Huber, M.~Misiak, M.~Niggetiedt, A.~Rehman et~al.,
  \emph{{The $Q_{1,2}${\textendash}$Q_7$ interference contributions to $b
  \rightarrow s \gamma $ at ${\mathcal O}(\alpha _{\mathrm s}^2)$ for the
  physical value of $m_c$}},
  \href{https://doi.org/10.1140/epjc/s10052-023-12270-8}{\emph{Eur. Phys. J. C}
  {\bfseries 83} (2023) 1108}
  [\href{https://arxiv.org/abs/2309.14707}{{\ttfamily 2309.14707}}].

\bibitem{Fael:2023gau}
M.~Fael, F.~Lange, K.~Sch{\"o}nwald and M.~Steinhauser, \emph{{Three-loop $b\to
  s\gamma$ vertex with current-current operators}},
  \href{https://doi.org/10.1007/JHEP11(2023)166}{\emph{JHEP} {\bfseries 11}
  (2023) 166} [\href{https://arxiv.org/abs/2309.14706}{{\ttfamily
  2309.14706}}].

\bibitem{Greub:2024mwp}
C.~Greub, H.M.~Asatrian, H.H.~Asatryan, L.~Born and J.~Eicher,
  \emph{{Three-loop contributions to b {\textrightarrow} s{\ensuremath{\gamma}}
  associated with the current-current operators}},
  \href{https://doi.org/10.1007/JHEP11(2024)058}{\emph{JHEP} {\bfseries 11}
  (2024) 058} [\href{https://arxiv.org/abs/2407.17270}{{\ttfamily
  2407.17270}}].

\bibitem{Misiak:2026sqy}
M.~Misiak et~al., \emph{{The inclusive $\bar B \to X_s \gamma$ decay rate with
  higher precision}},  \href{https://arxiv.org/abs/2608.29863}{{\ttfamily
  2608.29863}}.

\bibitem{Czaja:2026kop}
M.~Czaja, M.~Czakon, T.~Huber, M.~Misiak, M.~Niggetiedt, A.~Rehman et~al.,
  \emph{{NNLO QCD corrections to the weak radiative $B$-meson decay with exact
  dependence on $m_c$}},  \href{https://arxiv.org/abs/2608.29864}{{\ttfamily
  2608.29864}}.

\bibitem{Brune:2025zhd}
K.~Brune, T.~Huber and L.-T.~Moos, \emph{{Multi-parton contributions to
  $\overline{B}\to {X }_{s}\gamma $ at NLO}},
  \href{https://doi.org/10.1007/JHEP01(2026)142}{\emph{JHEP} {\bfseries 01}
  (2026) 142} [\href{https://arxiv.org/abs/2509.22564}{{\ttfamily
  2509.22564}}].

\bibitem{Chetyrkin:1996vx}
K.G.~Chetyrkin, M.~Misiak and M.~Munz, \emph{{Weak radiative B meson decay
  beyond leading logarithms}},
  \href{https://doi.org/10.1016/S0370-2693(97)00324-9}{\emph{Phys. Lett. B}
  {\bfseries 400} (1997) 206}
  [\href{https://arxiv.org/abs/hep-ph/9612313}{{\ttfamily hep-ph/9612313}}].

\bibitem{Kaminski:2012eb}
M.~Kaminski, M.~Misiak and M.~Poradzinski, \emph{{Tree-level contributions to
  $B \to X_s \gamma$}},
  \href{https://doi.org/10.1103/PhysRevD.86.094004}{\emph{Phys. Rev. D}
  {\bfseries 86} (2012) 094004}
  [\href{https://arxiv.org/abs/1209.0965}{{\ttfamily 1209.0965}}].

\bibitem{Huber:2014nna}
T.~Huber, M.~Poradzi{\'n}ski and J.~Virto, \emph{{Four-body contributions to $
  \overline{B}\to {X}_s\gamma $ at NLO}},
  \href{https://doi.org/10.1007/JHEP01(2015)115}{\emph{JHEP} {\bfseries 01}
  (2015) 115} [\href{https://arxiv.org/abs/1411.7677}{{\ttfamily 1411.7677}}].

\bibitem{Anastasiou:2002yz}
C.~Anastasiou and K.~Melnikov, \emph{{Higgs boson production at hadron
  colliders in NNLO QCD}},
  \href{https://doi.org/10.1016/S0550-3213(02)00837-4}{\emph{Nucl. Phys. B}
  {\bfseries 646} (2002) 220}
  [\href{https://arxiv.org/abs/hep-ph/0207004}{{\ttfamily hep-ph/0207004}}].

\bibitem{Chetyrkin:1981qh}
K.G.~Chetyrkin and F.V.~Tkachov, \emph{{Integration by parts: The algorithm to
  calculate $\beta$-functions in 4 loops}},
  \href{https://doi.org/10.1016/0550-3213(81)90199-1}{\emph{Nucl. Phys. B}
  {\bfseries 192} (1981) 159}.

\bibitem{Tkachov:1981wb}
F.V.~Tkachov, \emph{{A theorem on analytical calculability of 4-loop
  renormalization group functions}},
  \href{https://doi.org/10.1016/0370-2693(81)90288-4}{\emph{Phys. Lett. B}
  {\bfseries 100} (1981) 65}.

\bibitem{Laporta:2000dsw}
S.~Laporta, \emph{{High-precision calculation of multiloop Feynman integrals by
  difference equations}},
  \href{https://doi.org/10.1142/S0217751X00002159}{\emph{Int. J. Mod. Phys. A}
  {\bfseries 15} (2000) 5087}
  [\href{https://arxiv.org/abs/hep-ph/0102033}{{\ttfamily hep-ph/0102033}}].

\bibitem{Smirnov:2019qkx}
A.V.~Smirnov and F.S.~Chukharev, \emph{{FIRE6: Feynman Integral REduction with
  modular arithmetic}},
  \href{https://doi.org/10.1016/j.cpc.2019.106877}{\emph{Comput. Phys. Commun.}
  {\bfseries 247} (2020) 106877}
  [\href{https://arxiv.org/abs/1901.07808}{{\ttfamily 1901.07808}}].

\bibitem{Gehrmann-DeRidder:2003pne}
A.~Gehrmann-De~Ridder, T.~Gehrmann and G.~Heinrich, \emph{{Four particle phase
  space integrals in massless QCD}},
  \href{https://doi.org/10.1016/j.nuclphysb.2004.01.023}{\emph{Nucl. Phys. B}
  {\bfseries 682} (2004) 265}
  [\href{https://arxiv.org/abs/hep-ph/0311276}{{\ttfamily hep-ph/0311276}}].

\bibitem{Heinrich:2006sw}
G.~Heinrich, \emph{{Towards $e^+ e^- \to $ 3 jets at NNLO by sector
  decomposition}}, \href{https://doi.org/10.1140/epjc/s2006-02612-9}{\emph{Eur.
  Phys. J. C} {\bfseries 48} (2006) 25}
  [\href{https://arxiv.org/abs/hep-ph/0601062}{{\ttfamily hep-ph/0601062}}].

\bibitem{Kotikov:1990kg}
A.V.~Kotikov, \emph{{Differential equations method. New technique for massive
  Feynman diagram calculation}},
  \href{https://doi.org/10.1016/0370-2693(91)90413-K}{\emph{Phys. Lett. B}
  {\bfseries 254} (1991) 158}.

\bibitem{Remiddi:1997ny}
E.~Remiddi, \emph{{Differential equations for Feynman graph amplitudes}},
  \href{https://doi.org/10.1007/BF03185566}{\emph{Nuovo Cim. A} {\bfseries 110}
  (1997) 1435} [\href{https://arxiv.org/abs/hep-th/9711188}{{\ttfamily
  hep-th/9711188}}].

\bibitem{Argeri:2007up}
M.~Argeri and P.~Mastrolia, \emph{{Feynman Diagrams and Differential
  Equations}}, \href{https://doi.org/10.1142/S0217751X07037147}{\emph{Int. J.
  Mod. Phys. A} {\bfseries 22} (2007) 4375}
  [\href{https://arxiv.org/abs/0707.4037}{{\ttfamily 0707.4037}}].

\bibitem{Henn:2013pwa}
J.M.~Henn, \emph{{Multiloop integrals in dimensional regularization made
  simple}}, \href{https://doi.org/10.1103/PhysRevLett.110.251601}{\emph{Phys.
  Rev. Lett.} {\bfseries 110} (2013) 251601}
  [\href{https://arxiv.org/abs/1304.1806}{{\ttfamily 1304.1806}}].

\bibitem{Prausa:2017ltv}
M.~Prausa, \emph{{epsilon: A tool to find a canonical basis of master
  integrals}}, \href{https://doi.org/10.1016/j.cpc.2017.05.026}{\emph{Comput.
  Phys. Commun.} {\bfseries 219} (2017) 361}
  [\href{https://arxiv.org/abs/1701.00725}{{\ttfamily 1701.00725}}].

\bibitem{Huber:2005yg}
T.~Huber and D.~Ma{\^\i}tre, \emph{{HypExp, a Mathematica package for expanding
  hypergeometric functions around integer-valued parameters}},
  \href{https://doi.org/10.1016/j.cpc.2006.01.007}{\emph{Comput. Phys. Commun.}
  {\bfseries 175} (2006) 122}
  [\href{https://arxiv.org/abs/hep-ph/0507094}{{\ttfamily hep-ph/0507094}}].

\bibitem{Czakon:2005rk}
M.~Czakon, \emph{{Automatized analytic continuation of Mellin-Barnes
  integrals}}, \href{https://doi.org/10.1016/j.cpc.2006.07.002}{\emph{Comput.
  Phys. Commun.} {\bfseries 175} (2006) 559}
  [\href{https://arxiv.org/abs/hep-ph/0511200}{{\ttfamily hep-ph/0511200}}].

\bibitem{Gituliar:2015iyq}
O.~Gituliar, \emph{{Master integrals for splitting functions from differential
  equations in QCD}},
  \href{https://doi.org/10.1007/JHEP02(2016)017}{\emph{JHEP} {\bfseries 02}
  (2016) 017} [\href{https://arxiv.org/abs/1512.02045}{{\ttfamily
  1512.02045}}].

\bibitem{Huber:2005ig}
T.~Huber, E.~Lunghi, M.~Misiak and D.~Wyler, \emph{{Electromagnetic logarithms
  in $\bar B \to X_s l^+ l^-$}},
  \href{https://doi.org/10.1016/j.nuclphysb.2006.01.037}{\emph{Nucl. Phys. B}
  {\bfseries 740} (2006) 105}
  [\href{https://arxiv.org/abs/hep-ph/0512066}{{\ttfamily hep-ph/0512066}}].

\bibitem{Klappert:2020nbg}
J.~Klappert, F.~Lange, P.~Maierh{\"o}fer and J.~Usovitsch, \emph{{Integral
  reduction with Kira 2.0 and finite field methods}},
  \href{https://doi.org/10.1016/j.cpc.2021.108024}{\emph{Comput. Phys. Commun.}
  {\bfseries 266} (2021) 108024}
  [\href{https://arxiv.org/abs/2008.06494}{{\ttfamily 2008.06494}}].

\bibitem{Huber:2007dx}
T.~Huber and D.~Ma{\^\i}tre, \emph{{HypExp 2, Expanding hypergeometric
  functions about half-integer parameters}},
  \href{https://doi.org/10.1016/j.cpc.2007.12.008}{\emph{Comput. Phys. Commun.}
  {\bfseries 178} (2008) 755}
  [\href{https://arxiv.org/abs/0708.2443}{{\ttfamily 0708.2443}}].

\bibitem{Ferguson:1999pslq}
H.R.P.~Ferguson, D.H.~Bailey and S.~Arno, \emph{{Analysis of PSLQ, an integer
  relation finding algorithm}},
  \href{https://doi.org/https://doi.org/10.1090/S0025-5718-99-00995-3}{\emph{Math.
  Comp.} {\bfseries 68} (1999) 351}.

\bibitem{Ahnert:2011ode}
K.~Ahnert and M.~Mulansky, \emph{{Odeint - Solving Ordinary Differential
  Equations in C++}},
  \href{https://doi.org/http://dx.doi.org/10.1063/1.3637934}{\emph{AIP Conf.
  Proc. 1389} {\bfseries 1389} (2011) 1586}.

\bibitem{Hindmarsh:1983ode}
A.C.~Hindmarsh, \emph{{ODEPACK, A Systematized Collection of ODE Solvers}},
  {\emph{in: Scientific Computing, R. S. Stepleman et al. (eds.),
  North-Holland, Amsterdam, 1983 (vol. 1 of IMACS Transactions on Scientific
  Computation)} (1983) 55}.

\bibitem{Niggetiedt:974075}
M.~Niggetiedt, \emph{{Q}uark mass effects in form factors and hadronic {H}iggs
  production}, dissertation, RWTH Aachen University, Aachen, 2023.
\newblock 10.18154/RWTH-2023-11198.

\bibitem{Fael:2021kyg}
M.~Fael, F.~Lange, K.~Sch{\"o}nwald and M.~Steinhauser, \emph{{A semi-analytic
  method to compute Feynman integrals applied to four-loop corrections to the $
  \overline{\mathrm{MS}} $-pole quark mass relation}},
  \href{https://doi.org/10.1007/JHEP09(2021)152}{\emph{JHEP} {\bfseries 09}
  (2021) 152} [\href{https://arxiv.org/abs/2106.05296}{{\ttfamily
  2106.05296}}].

\bibitem{Liu:2022chg}
X.~Liu and Y.-Q.~Ma, \emph{{AMFlow: A Mathematica package for Feynman integrals
  computation via auxiliary mass flow}},
  \href{https://doi.org/10.1016/j.cpc.2022.108565}{\emph{Comput. Phys. Commun.}
  {\bfseries 283} (2023) 108565}
  [\href{https://arxiv.org/abs/2201.11669}{{\ttfamily 2201.11669}}].

\bibitem{Fael:2020tow}
M.~Fael, K.~Sch{\"o}nwald and M.~Steinhauser, \emph{{Third order corrections to
  the semileptonic b{\textrightarrow}c and the muon decays}},
  \href{https://doi.org/10.1103/PhysRevD.104.016003}{\emph{Phys. Rev. D}
  {\bfseries 104} (2021) 016003}
  [\href{https://arxiv.org/abs/2011.13654}{{\ttfamily 2011.13654}}].

\bibitem{Chen:2023osm}
L.~Chen, X.~Chen, X.~Guan and Y.-Q.~Ma, \emph{{Top-Quark Decay at
  Next-to-Next-to-Next-to-Leading Order in QCD}},
  \href{https://arxiv.org/abs/2309.01937}{{\ttfamily 2309.01937}}.

\bibitem{Chen:2023dsi}
L.-B.~Chen, H.T.~Li, Z.~Li, J.~Wang, Y.~Wang and Q.-f.~Wu, \emph{{Analytic
  third-order QCD corrections to top-quark and semileptonic b{\textrightarrow}u
  decays}}, \href{https://doi.org/10.1103/PhysRevD.109.L071503}{\emph{Phys.
  Rev. D} {\bfseries 109} (2024) L071503}
  [\href{https://arxiv.org/abs/2309.00762}{{\ttfamily 2309.00762}}].

\bibitem{Fael:2023tcv}
M.~Fael and J.~Usovitsch, \emph{{Third order correction to semileptonic $b\to
  u$ decay: Fermionic contributions}},
  \href{https://doi.org/10.1103/PhysRevD.108.114026}{\emph{Phys. Rev. D}
  {\bfseries 108} (2023) 114026}
  [\href{https://arxiv.org/abs/2310.03685}{{\ttfamily 2310.03685}}].

\bibitem{Fael:2020njb}
M.~Fael, K.~Sch{\"o}nwald and M.~Steinhauser, \emph{{Relation between the
  $\overline{\mathrm{MS}}$ and the kinetic mass of heavy quarks}},
  \href{https://doi.org/10.1103/PhysRevD.103.014005}{\emph{Phys. Rev. D}
  {\bfseries 103} (2021) 014005}
  [\href{https://arxiv.org/abs/2011.11655}{{\ttfamily 2011.11655}}].

\bibitem{Gunawardana:2019gep}
A.~Gunawardana and G.~Paz, \emph{{Reevaluating uncertainties in $\overline{B}
  \to X_{s}\gamma$ decay}},
  \href{https://doi.org/10.1007/JHEP11(2019)141}{\emph{JHEP} {\bfseries 11}
  (2019) 141} [\href{https://arxiv.org/abs/1908.02812}{{\ttfamily
  1908.02812}}].

\bibitem{Benzke:2020htm}
M.~Benzke and T.~Hurth, \emph{{Resolved $1/m_b$ contributions to $\bar B \to
  X_{s,d} \ell^+\ell^-$ and $\bar B \to X_s \gamma$}},
  \href{https://doi.org/10.1103/PhysRevD.102.114024}{\emph{Phys. Rev. D}
  {\bfseries 102} (2020) 114024}
  [\href{https://arxiv.org/abs/2006.00624}{{\ttfamily 2006.00624}}].

\bibitem{Bartocci:2024bbf}
R.~Bartocci, P.~B{\"o}er and T.~Hurth, \emph{{Renormalisation group evolution
  of the shape function g$_{17}$ in $ \overline{B}\to {X}_s\gamma $ and $
  \overline{B}\to {X}_s{\ell}^{+}{\ell}^{-} $ at subleading power}},
  \href{https://doi.org/10.1007/JHEP04(2025)066}{\emph{JHEP} {\bfseries 04}
  (2025) 066} [\href{https://arxiv.org/abs/2411.16634}{{\ttfamily
  2411.16634}}].

\bibitem{Benzke:2025ekp}
M.~Benzke, M.V.~Garzelli and T.~Hurth, \emph{{Update of the nonlocal
  sub-leading ${O}_1$ - ${O}_7$ contribution to $\bar B \to X_s \gamma$ at
  LO}}, \href{https://doi.org/10.1103/g6ds-ldx4}{\emph{Phys. Rev. D} {\bfseries
  113} (2026) 076006} [\href{https://arxiv.org/abs/2512.08902}{{\ttfamily
  2512.08902}}].

\bibitem{Fael:2026fxp}
M.~Fael, F.~Lange, K.~Sch{\"o}nwald and M.~Steinhauser, \emph{{The
  photon-energy spectrum in $B\to X_s\gamma$ to N$^3$LO: light-fermion and
  large-$N_{\textrm c}$ corrections}},
  \href{https://arxiv.org/abs/2603.15751}{{\ttfamily 2603.15751}}.

\bibitem{Hurth:2023paz}
T.~Hurth and R.~Szafron, \emph{{Refactorisation in subleading $\bar B \to X_s
  \gamma$}}, \href{https://doi.org/10.1016/j.nuclphysb.2023.116200}{\emph{Nucl.
  Phys. B} {\bfseries 991} (2023) 116200}
  [\href{https://arxiv.org/abs/2301.01739}{{\ttfamily 2301.01739}}].

\end{thebibliography}\endgroup

%\end{thebibliography}

\end{document}